\documentclass[aps,preprintnumbers,superscriptaddress,showpacs,twocolumn]{revtex4}
\usepackage{epsfig}
\usepackage{psfrag}
\usepackage{amsfonts}
\usepackage{graphicx}
\usepackage{dcolumn}
\usepackage{bm}

\begin{document}

\title{The refractive index in the viscous quark-gluon plasma}

\author{Bing-feng Jiang}
\email{jiangbf@iopp.ccnu.edu.cn} \affiliation{School of Mathematical and Physical Sciences, Hubei Institute
for Nationalities, Enshi, Hubei 445000,
 China }

\author{De-fu Hou}
\email{hdf@iopp.ccnu.edu.cn} \affiliation{Institute of Particle Physics, Huazhong Normal
University, Wuhan, Hubei 430079, China }

\author{Jia-rong Li}
\email{ljr@iopp.ccnu.edu.cn} \affiliation{Institute of Particle Physics, Huazhong Normal
University, Wuhan, Hubei 430079, China }
\author{Yan-jun Gao}
\email{gaoyjjwc@sina.com} \affiliation{School of Mathematical and Physical Sciences, Hubei Institute
for Nationalities, Enshi, Hubei 445000,
 China }

\date{\today}
\begin{abstract}
Under the framework of the viscous chromohydrodynamics, the gluon self-energy is derived for the quark-gluon plasma with shear viscosity. The viscous electric permittivity and magnetic permeability  are evaluated  from the gluon self-energy, through which the refraction index 
is investigated. The numerical analysis indicates that the refractive index becomes negative in some  frequency range. The start point for that frequency range is around the  electric permittivity pole, and the magnetic permeability pole determines the end point. As the increase of $\eta/s$, the frequency range for the negative refraction  becomes wider. 
\end{abstract}

\maketitle
\section{Introduction}

Quantum chromodynamics~(QCD) predicts that  deconfined phase  transition will take place at high temperature and/or high  density, as a result the nuclear matter will undergo a transition to quark-gluon plasma~(QGP). One main goal for Relativistic Heavy Ion Collider~(RHIC) and Large Hadronic Collider~(LHC) is to seek this new state of matter.  There are two striking findings at RHIC. One  is that the  deconfined hot medium behaviors as a nearly perfect fluid with a small viscosity\cite{result1,result2,result3,result4}. Several groups have applied viscous hydrodynamics to simulate the evolution of the produced matter in heavy ion collisions. The simulations successfully fit the observables at RHIC, such as the elliptic flow, the particle spectra, etc~\cite{liquid1,liquid3,liquid4,liquid5,liquid6,
liquid7,liquid8,liquid11,song}. The other is the strong jet quenching, which is believed to be a potential signal for the QGP formation\cite{wxn}.  It should be stressed that the first LHC results also strongly support   similar  conclusions as seen at RHIC\cite{alice,atlas}.

At the very early stage of the relativistic  heavy ion collisions, named glasma stage~\cite{glasma},  and the late stage of the evolution process in the near $T_c$ region in the so-called magnetic scenario for the QGP~\cite{monopole1, monopole2}, there are color-electric flux tubes which contain strong color-electric fields in them. Therefore, the color electromagnetic properties may play an important role in the evolution of  hot and dense matter produced in  heavy ion collisions. In addition, color electromagnetic properties could reflect  the response of QGP to external color current, so the study of them may be helpful for understanding the nature of QGP. However, to the best of our knowledge, the study on the color electromagnetic properties of  viscous QGP is scare in  literature. It makes sense to study how the viscosity affects them and how the viscous electromagnetic properties affect the evolution of produced matter in  heavy ion collisions.

Refraction index is one important electromagnetic property which reflects the  propagation of light in the medium. It can be determined  in terms of the electric permittivity $\varepsilon(\omega,k)$ and magnetic permeability $\mu_M(\omega,k)$.
In 1968, Veselago  proposed in theory that  the refraction index might be negative
in some special material\cite{veselago}.  That kind of medium is    in nature  consistent with the one proposed by Mandelstam in which the electromagnetic phase velocity propagates antiparallel to the energy flow\cite{agranovich}. But no any natural  material shows such special properties. Around 2000, by manipulating the array of small and closely spaced elements, scientists have constructed the negative refraction material in laboratory\cite{pendry,smith}, since then, the study on the negative refraction has attracted intensive interest. Recently, Amariti
et al. have studied the refraction index of the strong coupled system with the
string-inspired theory of AdS/CFT correspondence\cite{amariti}. Then, some
investigations have been carried out in strong coupled and correlation systems
along that line\cite{amariti2,ge,gao,amariti3}.
%
It is argued that the negative refraction  is a general phenomenon in some  frequency range in  charged fluid systems\cite{amariti3}. The probability  for the existence of negative refraction  in QGP is discussed as well in that literature\cite{amariti3}. Later, Juan Liu et al. extended the study of the refractive index of light to the weak coupled quark gluon system
within the framework of  the hard thermal loop perturbative theory\cite{wang}.

In this paper, we will make a first step to study the refraction index of gluon in the viscous QGP with the viscous chromohydrodynamics. Gluon is the QCD counterpart of photon. In addition, jet quenching has been proposed as a potential signal for the QGP and become an  active field in heavy ion collisions in last three decades, which is  relevant to the parton propagation in the hot medium. So the study of gluon refraction in QGP may be helpful for the understanding of the nature of  the QGP. According to the Refs.\cite{groot,arnold,teaney},
viscosity will modify the distribution functions of the constituents  of the QGP,
thus it will affect the gluon self-energy through which the electric permittivity and
magnetic permeability can  be derived.  Therefore,  viscosity will have an impact
on the refraction index.

It is argued  that chromohydrodynamics can describe
the polarization effect as the kinetic theory~\cite{manuel1}. In a recent
paper\cite{jiang2},  
some authors have extended the ideal chromohydrodynamics~\cite{manuel2,manuel3} to the viscous one in terms of the QGP kinetic theory and the distribution function modified by the shear viscosity. Under that framework, the polarization tensor is derived and the color-electric permittivity in the viscous QGP is studied in details\cite{jiang2}. Based on the color-electric permittivity, the induced color charge distribution\cite{jpg} and the corresponding wake potential\cite{npa} induced by the fast parton traveling through the viscous QGP have been investigated later.

In the present paper, by following the  gluon polarization tension derived from the viscous chromohydrodynamcs, we will derive  the  magnetic permeability and then study the refraction index in the QGP associated with shear viscosity. Our main results are as follows. Within the framework of the viscous chromohydrodynamics, the refraction index in the viscous QGP becomes negative in some frequency range. The start point of that frequency range is around the electric permittivity pole, while the magnetic permeability pole determines the  end point. In addition, with the increase of $\eta/s$, the frequency range for the existence of the negative refraction becomes broadening. 

The paper is organized as follows. In Section 2, we will briefly review the formulism of electromagnetic properties in medium. In Section 3, according to  the polarization tension derived from the viscous chromohydrodynamics, we evaluated the refraction index and discussed the viscous effect on it. Section 4 is summary and remarks.

The natural units $k_B=\hbar=c=1$, the metric $g_{\mu\nu}=(+,-,-,-)$ and  the following notations $K=(\omega,\textbf{k})$ are used in the paper.

\section{The electromagnetic properties in plasma}

In order to describe the electric and magnetic properties in plasma covariantly, it is convenient to introduce a pair of four-vectors $\widetilde{E}^\mu$, $\widetilde{B}^\mu$ in terms of the fluid four-velocity $u^\nu$
\begin{equation}\label{f1}
\widetilde{E}^\mu=u_\nu F^{\nu\mu}, \ \ \ \ \ \widetilde{B}^\mu=\frac{1}{2}\epsilon^{\mu\nu\lambda\rho}F_{\nu\lambda}u_\rho
\end{equation}
and
\begin{equation}\label{f2}
F^{\mu\nu}=u^\mu\widetilde{E}^\nu-\widetilde{E}^\mu u^\nu+\epsilon^{\mu\nu\lambda\rho}\widetilde{B}_\lambda u_\rho,
\end{equation}
where the Greek index $\mu$ is not confused with the magnetic permeability $\mu_M$ .
According to Eqs.(\ref{f1}) and (\ref{f2}),
one can obtain the Fourier-transformed free action
\begin{equation}\label{a1}
S_0=-\frac{1}{2} \int \frac{d^4K}{(2\pi)^4} \{\widetilde{E}^\mu(K)\widetilde{E}_\mu(-K)-\widetilde{B}^\mu(K)\widetilde{B}_\mu(-K)\}.
\end{equation}
Taking into account the interaction between the constituents of plasma, the correction to the action is
\begin{equation}\label{a2}
S_{int}=-\frac{1}{2} \int \frac{d^4K}{(2\pi)^4} A^\mu(-K)\Pi_{\mu\nu}(K)A^\nu(K),
\end{equation}
where $A^\mu(K)$ is vector boson field in momentum space, and $\Pi_{\mu\nu}(K)$ is polarization tensor which embodies the medium effects in plasma. In homogeneous and isotropic medium,the polarization tensor can be divided into longitudinal and transverse parts $\Pi_{\mu\nu}(K)=\Pi_L(K) P^L_{\mu\nu}(K)+\Pi_T(K) P^T_{\mu\nu}(K)$ with projector  defined  as $P^T_{00}=P^T_{0i}=P^T_{i0}=0$, $P^T_{ij}=\delta^{ij}-\frac{k^ik^j}{k^2}$, $P^L_{\mu\nu}=\frac{k^\mu k^\nu}{K^2}-g^{\mu\nu}-P^T_{\mu\nu}$\cite{kapusta,bellac}. Thus, the effective action including medium effects is
\begin{equation}\label{a3}
S_{eff}=S_0+S_{int},
\end{equation}
which  also can be described as
\begin{eqnarray}\label{a4}
S_{eff}&=&-\frac{1}{2} \int \frac{d^4K}{(2\pi)^4} [\varepsilon\widetilde{E}^\mu(K)\widetilde{E}_\mu(-K)\nonumber\\&-&
\frac{1}{\mu_M}\widetilde{B}^\mu(K)\widetilde{B}_\mu(-K)].
\end{eqnarray}
In (\ref{a4}), $\varepsilon$ and $\mu_M$ represent the electric permittivity and magnetic permeability respectively  which are the right quantities to describe the difference between the electric and magnetic properties of the vector field in the medium and those in the vacuum.
According to  Eqs.(\ref{a1}),(\ref{a2}) and (\ref{a4}), one can get the electric permittivity and magnetic permeability in plasma as following:
\begin{equation}\label{die}
\varepsilon(\omega,k)=1-\frac{\Pi_L(\omega,k)}{K^2}
\end{equation}
\begin{equation}\label{mag}
\frac{1}{\mu_M(\omega,k)}=1+\frac{K^2\Pi_T(\omega,k)-\omega^2\Pi_L(\omega,k)}{k^2K^2}
\end{equation}

We have briefly reviewed the electric and magnetic properties in homogeneous and isotropic plasma, the  detailed derivation also can be found in Refs.\cite{weldon,li,wang}. In addition, the extension of the discussion to the anisotropic medium was also addressed in Ref.\cite{wang}.

The refraction index is generally defined by the electric  permittivity and magnetic permeability as $n^2=\varepsilon(\omega,k)\mu_M(\omega,k)$ which is a square definition and not sensitive to  the simultaneous change of  signs of $\varepsilon$ and $\mu_M$. But it is proposed by Veselago that the simultaneous  change of positive  $\varepsilon$ and $\mu_M$ to negative $-\varepsilon$ and $-\mu_M$ corresponds to the transformation of the refraction index from one branch $n=\sqrt{\varepsilon(\omega,k)\mu_M(\omega,k)}$ to the other $n=-\sqrt{\varepsilon(\omega,k)\mu_M(\omega,k)}$, ie, the turn from the general refraction index to the negative  one. The physical nature of the negative refraction is that the electromagnetic phase velocity propagates opposite to the energy flow. For the detailed discussion on the physics of negative refraction, please refer to the Refs.\cite{veselago,agranovich,ramakrishna}.   The criterion for the negative refraction  is  $\varepsilon<0$ and $\mu<0$  simultaneously for real electric  permittivity and magnetic permeability medium.

If dissipation is taken into account, the situation is complicated. The electric  permittivity and magnetic permeability are generally complex-valued functions of $\omega$ and $k$, such as $\varepsilon(\omega,k)=\varepsilon_r(\omega,k)+i \varepsilon_i(\omega,k)$, $\mu_M(\omega,k)=\mu_r(\omega,k)+i \mu_i(\omega,k)$, so does the refraction index $n$. According to the phase velocity propagating antiparallel  to the energy flow, some authors have derived
the condition for  negative refraction in dissipative medium and found it is not necessary for   $\varepsilon_r<0$ and $\mu_r<0$ simultaneously\cite{mccall}. Later, another simple, convenient and widely adopted condition has been derived as \cite{depine}
\begin{equation}
n_{eff}=\varepsilon_r |\mu_M|+\mu_r|\varepsilon|<0,\label{neff}
\end{equation}
where $n_{eff}$ is called Depine-Lakhtakia index.
$n_{eff}<0$ implies ${\rm Re}\, n <0$, otherwise we will have a normal refraction index.  In this paper, we will use criterion (\ref{neff}) to study if there exists the negative refraction in the viscous QGP.

\section {The refraction index in the viscous quark-gluon plasma}

In this section, we will briefly review the derivation of the viscous chromohydrodynamics applicable to  QGP with shear viscosity. Then, we will solve fluid equations to obtain the gluon polarization tensor. According to the gluon self-energy, the electric permittivity and magnetic permeability are determined, through which the refraction index will be investigated.

\subsection{Kinetic theory}
The kinetic equations for quarks, antiquarks and gluons are given by~\cite{heinz1,heinz}
\begin{equation}
p^\mu D_\mu Q(p,x)+\frac{g}{2}p^\mu
\{F_{\mu\nu}(x),\partial^\nu_pQ(p,x)\}=C[Q,\bar{Q}, G],\label{kin1}
\end{equation}
\begin{equation}
p^\mu D_\mu \bar{Q}(p,x)-\frac{g}{2}p^\mu \{F_{\mu\nu}(x),
\partial^\nu_p\bar{Q}(p,x)\}=\bar{C}[Q,\bar{Q},G],\label{kin2}
\end{equation}
\begin{equation}
p^\mu \mathcal{D}_\mu G(p,x)+\frac{g}{2}p^\mu
\{\mathcal{F}_{\mu\nu}(x),
\partial^\nu_pG(p,x)\}=C_g[Q,\bar{Q},G].\label{kin3}
\end{equation}
$Q(p,x)$, $\bar{Q}(p,x)$ and $G(p,x)$ denote the
distribution functions of quark, antiquark and gluon respectively. $\partial_p^\nu$ represents
the four-momentum derivative and $\{ \cdots , \cdots\}$ is the anticommutator. $F_{\mu\nu}=\partial_\mu
A_\nu-\partial_\nu A_\mu-ig[A_\mu,A_\nu]$ represents the strength
tensor in the fundamental representation, and $\mathcal{F}_{\mu\nu}$
is its counterpart  in the adjoint representation.  $D_\mu$ and $\mathcal{D}_\mu$ represent the covariant derivatives
\begin{equation}
D_\mu=\partial_\mu-ig[A_\mu(x),\cdots], \ \ \ \ \
\mathcal{D}_\mu=\partial_\mu-ig[\mathcal{A}_\mu(x)\cdots].\label{cov}
\end{equation}
$A_\mu$ and $\mathcal{A}_\mu$  denote four-potentials in the
fundamental and adjoint representations respectively,
\begin{equation}
A_\mu(x)=A_{\mu,a}(x)\tau^a, \ \ \ \ \
\mathcal{A}_\mu(x)=T^aA_{\mu,a}(x),
\end{equation}
where $a=1,...,8$; $\tau^a$ and $T^a$ are the generators of group
SU(3) in the corresponding representations; $C$, $\bar{C}$ and $C_g$
denote the collision terms.

The transport equations are supplemented by the Yang-Mills equation,
\begin{equation}
D_\mu F^{\mu\nu}(x)=j^\nu(x), \label{yms}
\end{equation}
the color current $j^\nu(x)$  is given  in the fundamental
representation as
\begin{eqnarray}\label{c}
j^\nu(x)&=&-\frac{g}{2}\int_pp^\nu[Q(p,x)-\bar{Q}(p,x)-\frac{1}
{3}Tr[Q(p,x) \nonumber\\&-&\bar{Q}(p,x)]+2\tau^aTr[T^aG(p,x)]] \label{cl}
\end{eqnarray}
where $\int_p=\int \frac{d^4p}{(2\pi)^3}2 \Theta(p_0) \delta(p^2).$

Eqs.(\ref{kin1}),(\ref{kin2}),(\ref{kin3}), (\ref{yms}) and  (\ref{c}) make up the fundamental equations of the kinetic theory for the quark-gluon plasma. In the linear approximation of  QCD transport equation, by using the ideal, equilibrium distribution functions of constituents of the QGP, one can obtain the gluon self-energy\cite{bellac,heinz1,heinz}
\begin{equation}\label{htl1}
\Pi_L(\omega,k)=m_D^2(1-\frac{\omega^2}{k^2})
[1-\frac{\omega}{2k}\log[\frac{\omega+k}{\omega-k}]],
\end{equation}
 and
\begin{equation}\label{htl2}
\Pi_T(\omega,k)=\frac{1}{2}m_D^2[\frac{\omega^2}{k^2}+(1-\frac{\omega^2}{k^2})
\frac{\omega}{2k}\log[\frac{\omega+k}{\omega-k}]],
\end{equation}
where $m_D$ is the Debye  mass. (\ref{htl1}) and (\ref{htl2}) are consistent with those obtained in the hard thermal loop (HTL) approximation in diagrammatic methods at finite temperature field theory\cite{weldon,kapusta,bellac}. By combining with the HTL photon self-energy  and Eqs.(\ref{die}), (\ref{mag}) and (\ref{neff}), Juan Liu et al. have studied the refraction index of light in the QGP and found that it becomes negative in some frequency range.

\subsection{Viscous chromohydrodynamics}
In terms of Refs.\cite{groot,arnold,teaney}, viscosity will modified the distribution function of the constituents of QGP system. If only shear viscosity is taken into account,
the modified distribution function can be written as \cite{teaney}
\begin{equation}
Q=Q_o+\delta Q=Q_o+\frac{c'}{2T^3}\frac{\eta}{s}Q_o(1\pm Q_o)p^\mu
p^\nu\langle \nabla_\mu u_\nu \rangle. \label{dis}
\end{equation}
In Eq.(\ref{dis}), ``$+$'' is for boson, while ``$-$'' is for
fermion.
 $c'=\pi^4/90\zeta(5)$ and $c'=14
\pi^4/1350\zeta(5)$ are for massless boson~\cite{liquid11, teaney}
and massless fermion~\cite{dusling} respectively. $\langle
\nabla_\mu u_\nu \rangle = \nabla_\mu u_\nu + \nabla_\nu u_\mu -
\frac{2}{3} \Delta_{\mu\nu}\nabla_{\gamma}u^{\gamma}$, $\nabla_{\mu}
= (g_{\mu\nu} - u_{\mu}u_{\nu})\partial^{\nu}$, $\Delta^{\mu\nu}=
g^{\mu\nu} - u^{\mu}u^{\nu}$;   $\eta, s$,   $T$, $Q_o$
represent the  shear viscosity, the entropy density,  the temperature of the system and the ideal distribution function of boson or fermion.

It is very difficult to evaluate the gluon self-energy with the QGP kinetic theory
or finite temperature field theory associated with the  distribution functions modified by shear viscosity Eq.(\ref{dis}). Fortunately, the fluid equations are rather simpler than the kinetic theory and usually used to study the plasma properties.
By expanding the kinetic equations in  momenta moments  and truncating the
expansion at the second  level in terms of the ideal distribution function, chromohydrodynamics has been developed and been applied to study unstable modes of the QGP~\cite{manuel2,manuel3}.

By using the  quark distribution function modified by  shear viscosity Eq.(\ref{dis}) instead of the ideal one ($Q_o$) used in Refs.~\cite{manuel2,manuel3} and doing the same momentum moments in terms of collisionless version of kinetic equation (\ref{kin1}), we have extended the ideal chromohydrodynamic equations to the viscous
ones~\cite{jiang2}:
\begin{equation}
D_\mu n^\mu=0, \ \ \ \ \   D_\mu
T^{\mu\nu}-\frac{g}{2}\{F^\nu_\mu,n^\mu(x)\}=0   \label{con1}
\end{equation}
with
\begin{eqnarray}
 n^\mu(x)=\int_p  p^\mu Q(p,x), \
\    T^{\mu\nu}(x)=\int_p  p^\mu p^\nu Q(p,x). \label{co}
\end{eqnarray}

The four-flow $n^\mu$ and energy momentum tensor $T^{\mu\nu}$ can be
expressed in the form\cite{manuel3,jiang2}
\begin{eqnarray}
n^\mu=n(x) u^\mu, \ \ \ \ \ \ \ \ \ \ \ \ \ \ \ \ \ \  \ \ \ \ \ \ \ \nonumber\\
T^{\mu\nu}=\frac{1}{2}(\epsilon(x)+p(x))
\{u^\mu,u^\nu\}-p(x)g^{\mu\nu}+\pi^{\mu\nu}, \label{num}
\end{eqnarray}
where
\begin{eqnarray}
\pi^{\mu\nu}&=&\eta \langle \nabla^\mu u^\nu \rangle = \eta
\{(g^{\mu\rho}-u^\mu u^\rho)\partial_\rho u^\nu+(g^{\nu\rho}\nonumber\\&-&u^\nu
u^\rho)\partial_\rho u^\mu-\frac{2}{3}(g^{\mu\nu}-u^\mu
u^\nu)\partial_\sigma u^\sigma\}.\label{vis}
\end{eqnarray}

Because we only focus on the quark sector, the color
current  Eq.(\ref{c})  reads
\begin{equation}
j^\mu(x)=-\frac{g}{2}(nu^\mu-\frac{1}{3}Tr[nu^\mu]).\label{curr}
\end{equation}
Eqs.(\ref{con1}),(\ref{num}) and (\ref{curr}) make up the basic set of equations of the viscous chromohydrodynamics.
In those equations,  $n$, $\epsilon$ and $p$ represent  the particle density, the energy density and pressure respectively. These quantities are $N_c\times N_c$ matrices
in color space\cite{manuel3}. If $\eta=0$, the distribution function
remains the ideal form, $\pi^{\mu\nu}$ will be absent in (\ref{num})
and the chromohydrodynamic equations will turn to the ideal ones\cite{manuel3}.

\subsection{Gluon self-energy}

Linearizing the hydrodynamic quantities around the stationary,
colorless and homogeneous state which is described by
$\bar{n}$,$\bar{u}^\mu$,$\bar{p}$ and $\bar{\epsilon}$, as an
example, the particle density is written as
\begin{equation}\label{linearization}
n(x)=\bar{n}+\delta n(x).
\end{equation}
The stationary and fluctuation quantities satisfy
$\delta n\ll \bar{n}$ and $D_\mu\bar{n}=0$. The corresponding parts
of other hydrodynamic quantities have the similar properties. The
color current $j^\mu(x)$ vanishes in the stationary state. All the
fluctuations of the hydrodynamics quantities can contain both
colorless and colored components, for example,
\begin{equation}
\delta n=\delta n_0 I_{\alpha\beta}+\frac{1}{2}\delta n_a
\tau^a_{\alpha\beta},
\end{equation}
where $\alpha,\beta=1,2,3$ are color indices and $I$ is the identity
matrix \cite{manuel3}.

Substituting the linearized hydrodynamic quantities like
Eq.(\ref{linearization}) into Eq.(\ref{num}) and their corresponding
conservation equations (\ref{con1}) and projecting them on
$\bar{u}^\mu$ and
$(g^{\mu\nu}-\bar{u}^\mu\bar{u}^\nu)$,
then, considering only the equations for colored parts of fluctuations and performing
the Fourier transformation, one can obtain equations which can
describe color phenomena in the viscous QGP\cite{jiang2}
\begin{equation}
\bar{n}k_\mu\delta u^\mu_a+ k_\mu\delta
n_a\bar{u}^\mu=0,\label{con1f}
\end{equation}
\begin{equation}
\bar{u}^\mu
k_\mu\delta\epsilon_a+(\bar{\epsilon}+\bar{p})k_\mu\delta
u^\mu_a=0,\label{con21f}
\end{equation}
\begin{eqnarray}
(\bar{\epsilon}+\bar{p})(\bar{u}\cdot K)\delta
u^\nu_a+(-k^\nu+\bar{u}^\nu(\bar{u}\cdot K))\delta p_a \nonumber\\+ \eta\{(K^2
-(K\cdot \bar{u}))\delta u^\nu_a  + (k^\mu k^\nu-k^\mu
\bar{u}^\nu)\delta u_{\mu,a}\nonumber\\+\frac{2}{3}(\bar{u}^\nu (K\cdot
\bar{u})-k^\nu)k_\rho\delta u^\rho_a\}=ig\bar{n}\bar{u}_\mu
F_a^{\mu\nu}(K).\label{con22f}
\end{eqnarray}

We introduce an EoS $\delta p_a=c_s^2\delta\epsilon_a$ to complete the fluid equations, the explicit formulism for $c_s$ will be introduced later. According to Eqs.(\ref{con1f}),(\ref{con21f}), (\ref{con22f}) and the introduced EoS,
we can obtain the colored fluctuations of
hydrodynamic quantities $\delta n_a$, $\delta u_{\nu,a}$ and
$\delta\epsilon_a$.
Due to the color fluctuations of the hydrodynamic quantities, the
color current fluctuation is given by
\begin{equation}
\delta j^\mu_a=-\frac{g}{2}(\bar{n}\delta u^\nu_a+ \delta
n_a\bar{u}^\mu-\frac{1}{3}Tr[\bar{n}\delta u^\mu_a+ \delta
n_a\bar{u}^\mu]).
\end{equation}
Substituting into the solved $\delta n_a$ and $\delta u^\mu_a$,
according to the relation between the current and the gauge field in
the linear response theory $\delta
j^\mu_a(K)=-\Pi^{\mu\nu}_{ab}(K)A_{\nu,b}(K)$, one can abstract the
polarization tensor $\Pi^{\mu\nu}_{ab}(K)$\cite{jiang2}
\begin{eqnarray}
\Pi_{ab}^{\mu\nu}(\omega,k)&=&-\delta_{ab} \{\omega_p^2
 \cdot\frac{1}{1+D(K^2-(K\cdot \bar{u})^2)}\cdot\frac{1}{(K\cdot
\bar{u})^2}\nonumber\\&\cdot&[(K\cdot \bar{u})(\bar{u}^\mu k^\nu+k^\mu
\bar{u}^\nu)-K^2\bar{u}^\mu \bar{u}^\nu\nonumber\\&-&(K\cdot
\bar{u})^2g^{\mu\nu}+(B+E)\cdot[K^2(K\cdot \bar{u})(\bar{u}^\mu
k^\nu\nonumber\\&+&k^\mu\bar{u}^\nu)-k^\mu k^\nu (K\cdot
\bar{u})^2-K^4\bar{u}^\mu \bar{u}^\nu]]\},\label{pol}
\end{eqnarray}
where $\omega_p$ is the  plasma frequency and
\begin{eqnarray}
 B&=&-\frac{c_s^2}
  {\omega^2-c_s^2k^2},\ \ \ \ \ \  \ \ \ \ \ \  D=\frac{\eta}{sT\omega},\nonumber\\
  E
 &=&-\frac{\frac{\eta\omega}{sT}
 (1+4\frac{c_s^2k^2}
  {\omega^2-c_s^2k^2})}
  {3\omega^2-3c_s^2k^2-
 4\frac{\eta\omega k^2}
{sT}}.\label{d}
\end{eqnarray}
It is easy to test that $\Pi_{ab}^{\mu\nu}=\Pi_{ab}^{\nu\mu}$ and $k_\mu \Pi_{ab}^{\mu\nu}=0$. In the further considerations, we suppress the color indices $a,b$.

According to the projector, we can obtain the longitudinal and transverse gluon self-energy
\begin{eqnarray}
\Pi_L(K)=\frac{K^2}{k^2}\Pi^{00}(K),\label{pil}\\
\Pi_T(K)=\frac{1}{2}(\delta_{ij}-
\hat{k}_i\hat{k}_j)\Pi^{ij}(K),\label{pit}
\end{eqnarray}
with $\hat{k}_i=k_i/k$.

\begin{figure}
\begin{minipage}[h]{0.48\textwidth}
\centering{\includegraphics[width=7.5cm,height=4.5cm]{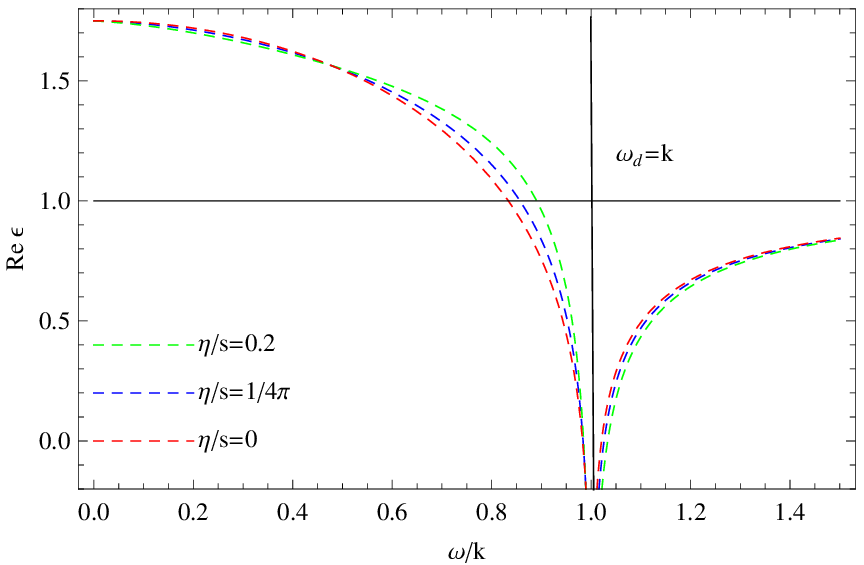}}
\end{minipage}
\begin{minipage}[h]{0.48\textwidth}
\centering{\includegraphics[width=7.5cm,height=4.5cm] {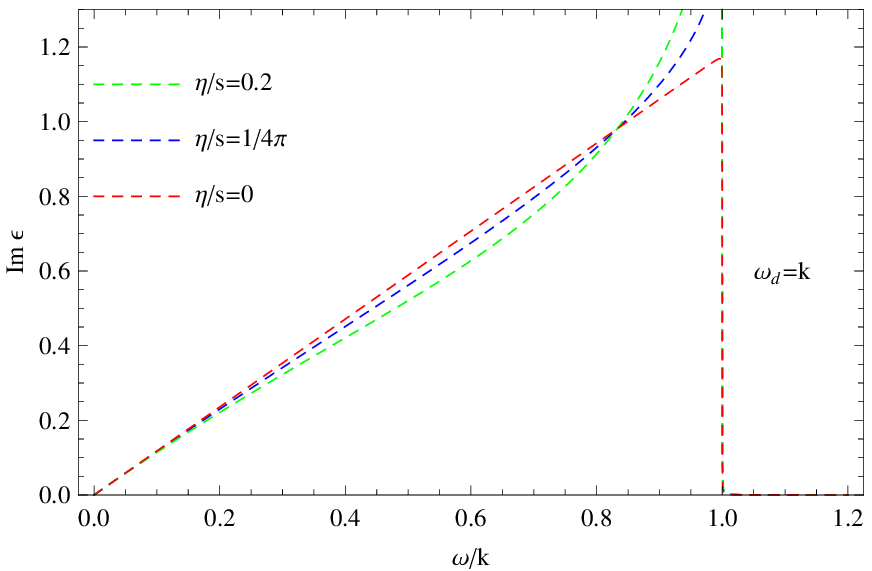}}
\end{minipage}
\caption{(color online) The electric permittivity in the viscous QGP.  Top panel: the real part. Bottom panel: the imaginary part. The dashed red, blue and green curves are for the cases of $\eta/s=0, 1/4\pi, 0.2$ respectively.} \label{electric}
\end{figure}

We have briefly reviewed the determination  of the gluon self-energy in the QGP associated with shear viscosity with the viscous chromohydrodynamic approach. For details  please refer to Ref.\cite{jiang2}. Through the derivation, shear viscosity is encoded in the gluon self-energy. Combining with Eqs.~(\ref{pil}), (\ref{pit}), (\ref{die}), (\ref{mag}) and (\ref{neff}), we can study the refraction index of gluon in the viscous QGP.

\subsection{Numerical results}

Before we do a further analysis on the refraction index,  we should
determine the sound speed $c_s$ first. Mannarelli and  Manuel have investigated
collective unstable modes of QGP with  the  ideal chromohydrodynamic
approach~\cite{manuel3} as well as the kinetic theory~\cite{manuel1}. They found that when one uses ``the effective speed of sound''
$c_s=\sqrt{\frac{1}{3(1+\frac{1}{2y}\log\frac{1-y}{1+y})}+\frac{1}{y^2}}$
$(y=\frac{k}{\omega})$, the results in the  chromohydrodynamic approach agree well with those in the kinetic theory  in the same
setting (see discussion in Appendix in Ref.~\cite{manuel1}). In this
paper, we also use the effective speed of sound.
\begin{figure}
\begin{minipage}[h]{0.48\textwidth}
\centering{\includegraphics[width=7.5cm,height=4.635cm] {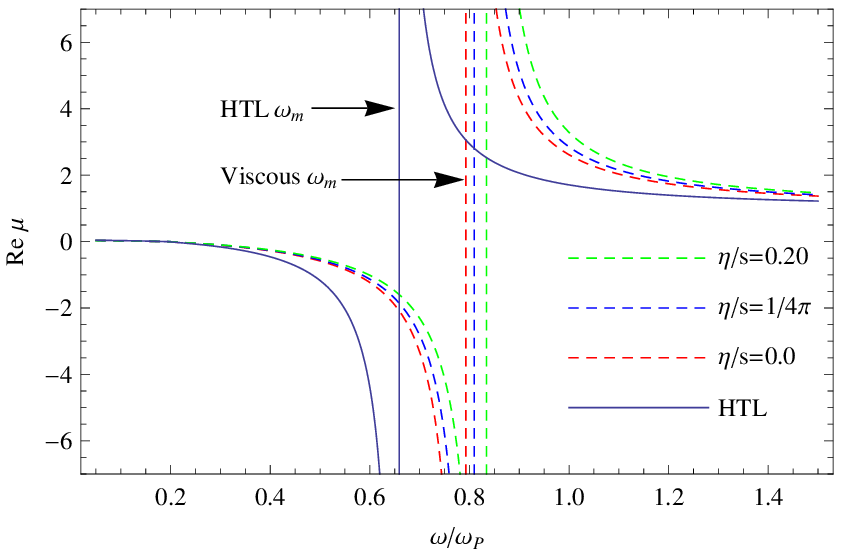}}
\end{minipage}
\begin{minipage}[h]{0.48\textwidth}
\centering{\includegraphics[width=7.5cm,height=4.5cm] {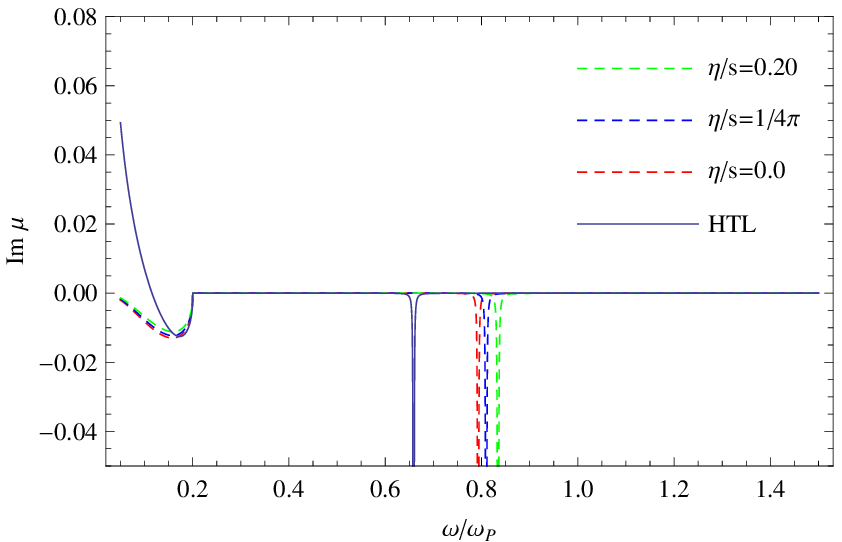}}
\end{minipage}
\caption{(color online) The magnetic permeability in the QGP.  Top panel: the real part. Bottom panel: the imaginary part. The dashed red, blue and green curves are for the cases of $\eta/s=0, 1/4\pi, 0.2$ respectively, while the solid blue curve is for the HTL case. } \label{magnetic}
\end{figure}

In this paper, we can not determine shear viscosity coefficient itself in viscous chromohydrodynamics, but regard it as an input parameter to study the viscous effect on the refraction index of the QGP. A small value of the ratio for shear viscosity to entropy density $\eta/s\leq 0.2$ has been deduced from comparison of casual viscous hydrodynamic simulation results with the RHIC data\cite{song}, which is less than three times of the famous bound result $\eta/s=\frac{1}{4\pi}$ of the strongly couple conformal field theory determined by the AdS/CFT correspondence\cite{bound}. Numerical results of the refraction index are presented with those explicit values of $\eta/s$. In addition, in numerical analysis, such scales $k=0.2\omega_p$ and $T=\omega_p$ are used to study the $\omega$-dependent behavior of the refraction index.

In terms of the relation
between Eq.~(\ref{pil}) and (\ref{die}),  one can
obtain the electric permittivity in soft momentum approximation~\cite{jiang2}
\begin{eqnarray}
 \varepsilon(\omega,k)
&=&1+\frac{3\omega_P^2}{k^2}[1-\frac{\omega}{2k}
 (\log|\frac{\omega+k}{\omega-k}|-i\pi\Theta(k^2-\omega^2))]
\nonumber\\&-&\frac{12\omega_P^2}
 {k^2}\frac{\eta\omega}{sT}\times
 \{1-\frac{\omega}{k}
 \log|\frac{\omega+k}{\omega-k}|
 \nonumber\\&+&\frac{\omega^2}{4k^2}
 (\log|\frac{\omega+k}{\omega-k}|)^2-
 \frac{\omega^2}{4k^2}\pi^2\Theta(k^2-\omega^2)\nonumber\\&+& i
 (\frac{\omega}{k}\pi-\frac{\omega^2}{2k^2}\pi
 \log|\frac{\omega+k}{\omega-k}|)\Theta(k^2-\omega^2)\}, \label{vdief}
\end{eqnarray}
where $\Theta$ is the step function. It should be noted that the same result  has been obtained  in nonlinear viscous chromohydrodynamics in the soft limit of $\omega,k\ll T$   in a recent literature\cite{ramos}.
Here, we plot the real and imaginary parts of the  electric permittivity in the viscous QGP in Fig.1. The main findings are as following. First, there is a frequency pole for the real part at $\omega_{d}=k$, which is  just the inflexion of the imaginary part. Second, if such relation $m^2_D=3\omega^2_P$ is adopted\cite{mustafa},  when $\eta/s=0$, $\varepsilon(\omega,k)$ recovers the HTL result obtained by the kinetic theory or finite temperature field theory\cite{weldon,kapusta,bellac,heinz1,heinz}. Third, the viscous corrections to both the real and imaginary parts of $\varepsilon(\omega,k)$ are small. For detailed discussion, please refer to Ref.\cite{jiang2}.

According to the relation between $\Pi_L$, $\Pi_T$ and $\mu_M(\omega,k)$, the magnetic permeability in the viscous QGP can be derived from Eqs.~(\ref{pil}), (\ref{pit}) and (\ref{mag})
\begin{eqnarray}
\mu_M(\omega,k)
=\frac{1}{1+\frac{\omega_p^2}{k^2}\cdot\frac{1}{1-\frac{\eta}{s}\cdot\frac{k^2}{T\omega}}
+\frac{\omega^2}{k^2}\cdot(\varepsilon(\omega,k)-1)}.\label{vmag}
\end{eqnarray}
We present the real and imaginary parts of magnetic permeability of the viscous QGP
in top and bottom panels respectively in Fig.2. For comparison, we also display the HTL results  as well. The dashed red, blue and green curves are for the viscous cases of $\eta/s=0, 1/4\pi, 0.2$ respectively, while the blue solid curves are for the HTL results.
Both  the real and imaginary parts of magnetic
permeability show up a frequency pole  $\omega_m$. Its position
is around $0.65\omega_P$ for the HTL results, but around $0.8\omega_P$ for the viscous cases.
In addition, it is easy to see that the  frequency pole  shifts to large frequency region with the increase of $\eta/s$.

From Eqs.(\ref{vdief}), (\ref{vmag}) and (\ref{neff}), we can determine the
Depine-Lakhtakia index in the viscous QGP. We display its numerical results with different values of $\eta/s$ in Fig.3 as well as the HTL result evaluated from Eqs.(\ref{htl1}), (\ref{htl2}), (\ref{die}), (\ref{mag}) and (\ref{neff}). As shown in Fig.3, there is a quite large frequency range for $n_{eff}<0$. In terms of the discussion in the section 2, in that frequency range refraction index becomes negative, ie. ${\rm Re}\, n<0$. With the increase of $\eta/s$, the frequency range for the negative refraction becomes wider. In addition, the frequency range for negative refraction in the viscous QGP is much wider than that of the  HTL case.

In Fig.3, one can see that, with the increase of the frequency, there is an inflexion for $n_{eff}$ where $n_{eff}$ changes from positive to negative value when the frequency
$\omega$ is around $\omega_d$, and
$n_{eff}$ is negative until $\omega=\omega_m$. The numerical analysis in Fig.1, Fig.2 and Fig.3 shows that the start point of the frequency region for negative refraction is around the electric permittivity $\omega_d$,
while the  magnetic permeability pole $\omega_m$ determines the end point. Note that the electric permittivity poles in the viscous cases superpose each other, whose position   coincides with that of the HTL result, as shown in Fig.1. Therefore, the start points for negative refraction show no appreciable distinction   among all curves in Fig.3.  From Fig.2, one can see that the magnetic permeability pole shifts to large frequency region with the increase of $\eta/s$, which leads to  an enlargement of the frequency range for negative refraction. The magnetic permeability pole is around $0.65\omega_P$ for the HTL result, but around $0.8\omega_P$ for the viscous cases, which results in the fact that the frequency range for negative refraction in the viscous QGP is much wider than that of the  HTL case.

\begin{figure}
\includegraphics[width=7.5cm,height=4.635cm] {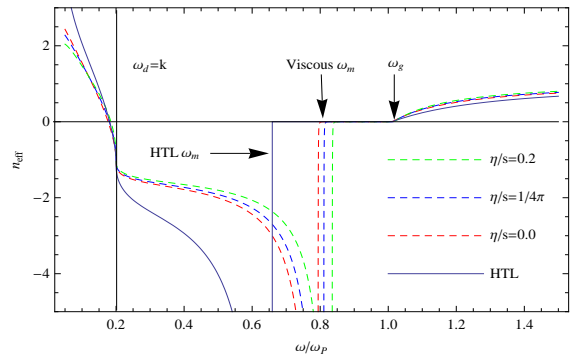}
\caption{(color online) The Depine-Lakhtakia index for the QGP. The dashed red, blue and green curves are for the cases of $\eta/s=0, 1/4\pi, 0.2$ respectively, while the solid blue curve is for the HTL case.} \label{effn}
\end{figure}


\begin{figure}
\begin{minipage}[h]{0.48\textwidth}
\centering{\includegraphics[width=7.5cm,height=4.635cm] {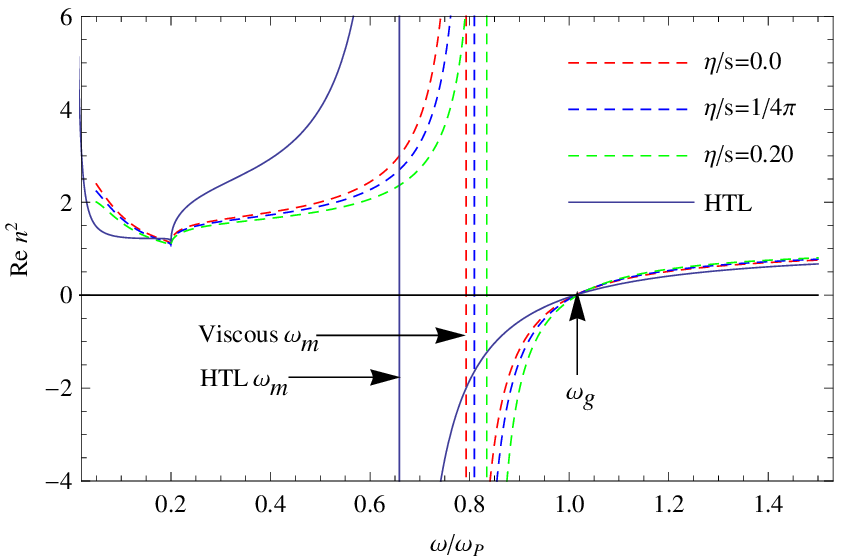}}
\end{minipage}
\begin{minipage}[h]{0.48\textwidth}
\centering{\includegraphics[width=7.5cm,height=4.5cm] {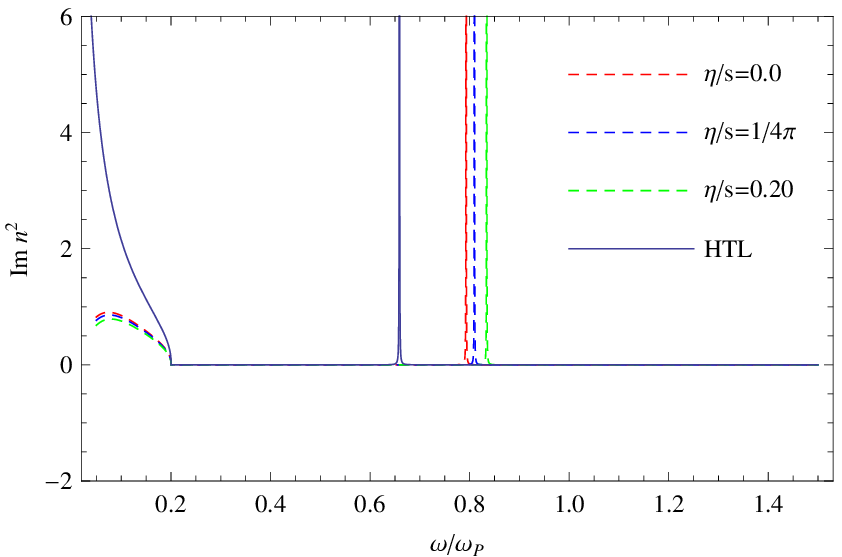}}
\end{minipage}
\caption{(color online) The real and imaginary parts of $n^2$ in the  QGP. Top panel: the real part. Bottom panel:imaginary part. The dashed red, blue and green curves are for the cases of $\eta/s=0, 1/4\pi, 0.2$ respectively, while the solid blue curve is for the HTL case.} \label{fig3}
\end{figure}


From Fig.3, it is shown that both viscous curves and the HTL curve intersect  one point at $\omega=\omega_g$. When $\omega>\omega_g$, $n_{eff}>0$ for all curves, which implies a general refraction index. A frequency gap $\omega\in[\omega_m,\omega_g]$ is illustrated for $n_{eff}=0$, in which $n^2<0$ as shown in Fig.4.
It is argued that that result has not been  reported in earlier literature\cite{wang}.
The light does not propagate in that frequency gap, 
because 
the refraction index is pure  imaginary and the electromagnetic wave is damped severely \cite{wang}. 

\begin{figure}
\begin{minipage}[h]{0.48\textwidth}
\centering{\includegraphics[width=7.5cm,height=4.635cm] {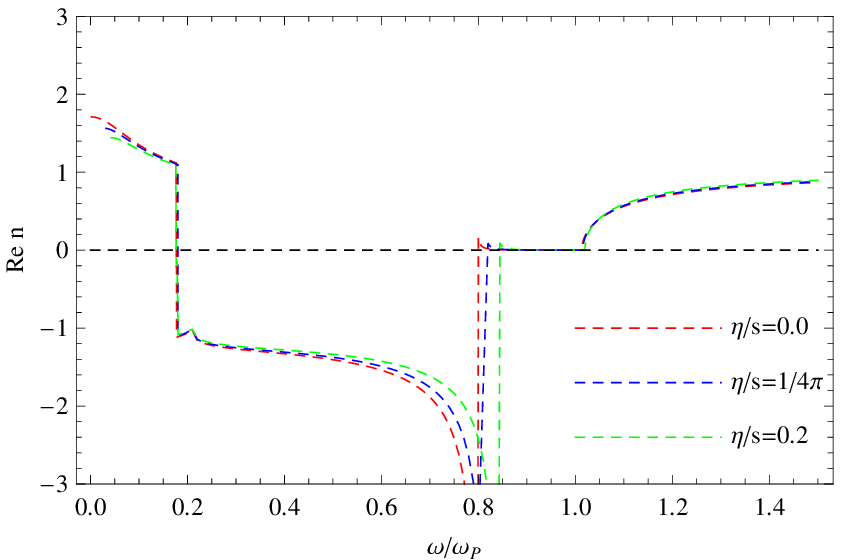}}
\end{minipage}
\begin{minipage}[h]{0.48\textwidth}
\centering{\includegraphics[width=7.5cm,height=4.5cm] {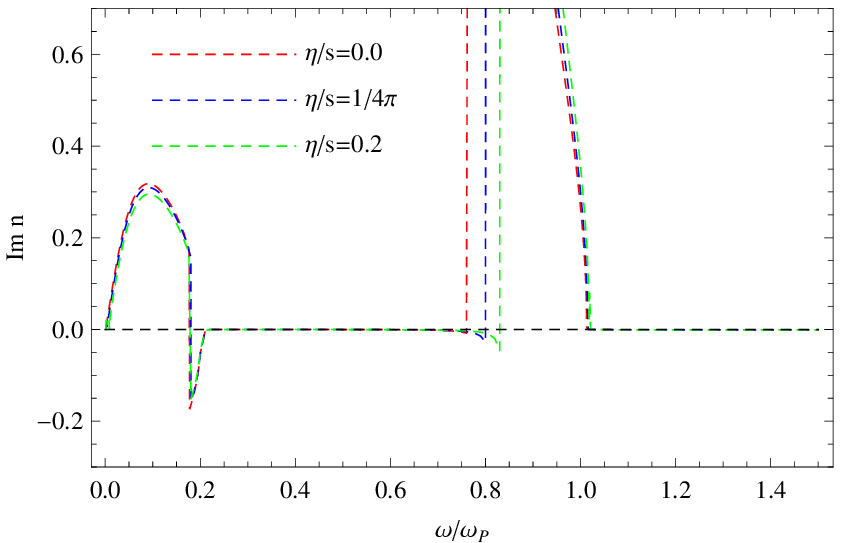}}
\end{minipage}
\caption{(color online) The refraction index in the viscous QGP, Top panel: the real part. Bottom panel:imaginary part. The dashed red, blue and green curves are for the cases of $\eta/s=0, 1/4\pi, 0.2$ respectively.} \label{fig3}
\end{figure}

To obtain  the refraction index $n$, one has to study the square root of  $n^2=\varepsilon\mu_M$. The complex value number $n^2=\varepsilon\mu_M$ possesses two square roots,
\begin{equation}\label{square}
n=\pm \sqrt{|n^2|} e^{i (\phi/2)}.
\end{equation}
In (\ref{square}), $\phi$ is the argument of  $n^2=\varepsilon\mu_M$ which can be expressed as
\begin{equation}
\phi= \phi_\varepsilon + \phi_\mu,
\end{equation}
where  $\phi_\varepsilon=\varepsilon_i/\varepsilon_r$ and $\phi_\mu=\mu_i/\mu_r$ are the arguments of  $\varepsilon$ and $\mu_M$ respectively. But which root in Eq.(\ref{square}) is to be chosen? In terms of the criterion (\ref{neff}) that $n_{eff}<0$ implying ${\rm Re}\, n <0$, otherwise  ${\rm Re}\, n>0$, and the arguments of $\varepsilon$ (\ref{vdief}) and $\mu_M$ (\ref{vmag}), we present the real and imaginary parts of $n$ in the QGP with different $\eta/s$ values in Fig.5. One can see that the properties of  the real part of $n$ is qualitatively consistent with the corresponding ones of $n_{eff}$. Only the frequency is around the magnetic permeability pole, the obvious difference demonstrate for the real and imaginary parts of $n$ with the different $\eta/s$. Otherwise, the viscous correction to them is very trivial. The HTL results of $n$ is also demonstrated in Ref.\cite{wang}.

Due to simplicity and applicability to describe polarization effect, we  have applied the viscous chromohydrodynamics, which is derived from the QGP kinetic theory and the viscosity-modified distribution function, to determine the polarization tensor and investigate the refraction index in the viscous QGP. It should be noted that some dynamical information will be lost during the derivation from the kinetic theory to the chromohydrodynamics\cite{manuel1,manuel3,ramos}. However, in many cases the discrepancies between both approaches of the kinetic theory and the chromohydrodynamics can be alleviated by using effective parameters as inputs in the hydrodynamic formalisms\cite{manuel1,ramos}. Nevertheless, such phenomenological model of the chromohydrodynamics could capture the some correct physics of the QGP\cite{manuel3,ramos}. In view of the difficulty in investigating  the viscous effect on the color electromagnetic properties of QGP in microscopic kinetic theory description, we expect that we could obtain some insight on the physics of the  problem by applied the viscous chromohydrodynamics.


\section{Summary}
\label{summary}

In this paper, within the framework of the viscous chromohydrodynamics, the gluon self-energy has been evaluated in the QGP associated with shear viscosity, through which the electric permittivity and magnetic permeability have been derived. Based on the viscous $\varepsilon(\omega,k)$ and $\mu_M(\omega,k)$, we have investigated the Depine-Lakhtakia index $n_{eff}$ and the refraction index $n$ in the viscous QGP. For comparison, we have also presented the corresponding HTL reslults. $n_{eff}<0$ implies  the negative real part of $n$, which signifies the negative refraction in the medium. The numerical analysis shows that
i) the refraction index becomes negative in some frequency range;
ii) the start point of that frequency range is around the pole of electric permittivity $\varepsilon(\omega,k)$, and the magnetic permeability pole determines the end point;
iii) with the increase of the $\eta/s$, the frequency range for  the negative refraction becomes broader.  In addition, the frequency range for negative refraction in the viscous chromohydrodynamics is wider than that of the HTL perturbation theory. The numerical analysis also indicates that viscous properties of poles for $\varepsilon$ and $\mu_M$ are responsible for that difference. 

The criterion (\ref{neff}) $n_{eff}<0$ has been widely used to judge the existence of the negative refraction in a medium. It is  interesting to study the interplay between the  modes propagation (${\rm Re}\, n$)  and dissipation (${\rm Im}\, n$) in the frequency region for negative refraction.  For instance ,in strongly coupled system with the framework of AdS/CFT correspondence,it was found that there exists strong dissipation in the frequency region for negative refraction, only around $\omega \rightarrow 0$,  propagation may dominate over dissipation\cite{amariti}. How  the shear viscosity impacts the  mode propagation in QGP deserves further comprehensive studies.

{\bf Acknowledgment} We are very
grateful to  A.Amariti,  M.J. Luo, H.C.Ren , Q. Wang and P.F. Zhuang for helpful discussion.
This work is supported partly by National Natural Science Foundation of
China under Grant Nos.\ 11147012, 11275082, 10975060, 11135011, 11221504 and Foundation of Hubei Institute for Nationalities under Grant Nos.\ MY2011B002,4129030.


\end{document}